\documentclass{article}
\usepackage{spconf,amsmath,graphicx}
\usepackage{hyperref}
\usepackage{booktabs}
\usepackage{multirow}
\usepackage{subcaption}
\usepackage{makecell}
\usepackage{float}
\usepackage{placeins}
\usepackage{textcomp}
\usepackage{comment}
\usepackage{enumitem}

\usepackage[font=footnotesize, skip=1pt]{caption}
\title{Learning Perceptual Representations for Gaming NR-VQA with Multi-Task FR Signals}
\name{\shortstack{%
Yu-Chih Chen\textsuperscript{1,2}, Michael Wang\textsuperscript{1}, Chieh-Dun Wen\textsuperscript{1}, Kai-Siang Ma\textsuperscript{1}\\
Avinab Saha\textsuperscript{2}\textsuperscript{\textdagger}\thanks{{\textdagger}Work done at UT Austin; Avinab Saha is now at Google Research.}, Li-Heng Chen\textsuperscript{2,3}, Alan Bovik\textsuperscript{2}%
}}
\address{\textsuperscript{1}National Yang Ming Chiao Tung University,\textsuperscript{2}The University of Texas at Austin,
\textsuperscript{3}Netflix Inc.}

\begin{document}
\maketitle
\begingroup
\renewcommand\thefootnote{}
\footnotetext{This work was supported by the National Science and Technology Council, Taiwan, under Grant NSTC 115-2222-E-A49-005-MY2.}
\endgroup
\begin{abstract}
No-reference video quality assessment (NR-VQA) for gaming videos is challenging due to limited human-rated datasets and unique content characteristics including fast motion, stylized graphics, and compression artifacts. We present MTL-VQA, a multi-task learning framework that uses full-reference (FR) quality metrics as supervisory signals to learn perceptually meaningful features without human labels during pretraining. By jointly optimizing multiple complementary proxy FR objectives with adaptive task weighting, our approach learns shared representations that transfer effectively to downstream NR-VQA. Experiments on gaming video datasets show that MTL-VQA achieves competitive performance against state-of-the-art methods in both mean opinion score-supervised and label-efficient or self-supervised settings.
\end{abstract}
\begin{keywords}
no-reference video quality assessment, multi-task learning, representation learning
\end{keywords}

\vspace{-4mm}
\section{Introduction}
\label{sec:intro}
\vspace{-3mm}
Cloud gaming has grown rapidly with high-bandwidth wireless networks and modern mobile devices. Unlike natural videos, gaming content is computer-generated and exhibits distinctive statistics—including fast motion, stylized graphics, user-interface overlays (HUD), and codec-induced artifacts—that violate assumptions commonly exploited by natural-scene-based image quality assessment (IQA) and video quality assessment (VQA) models. In the cloud gaming setting, reference videos are not accessible on the client side; hence no-reference (NR) VQA is required at inference to monitor quality of experience under bitrate, latency, and computational constraints. However, NR-VQA is intrinsically more challenging than full-reference (FR) VQA because the model must infer perceptual quality without access to the original reference video. In this setting, content and distortion cues are entangled, making robust quality prediction substantially harder.

\begin{figure}[!ht]
  \centering
  \includegraphics[width=240pt]{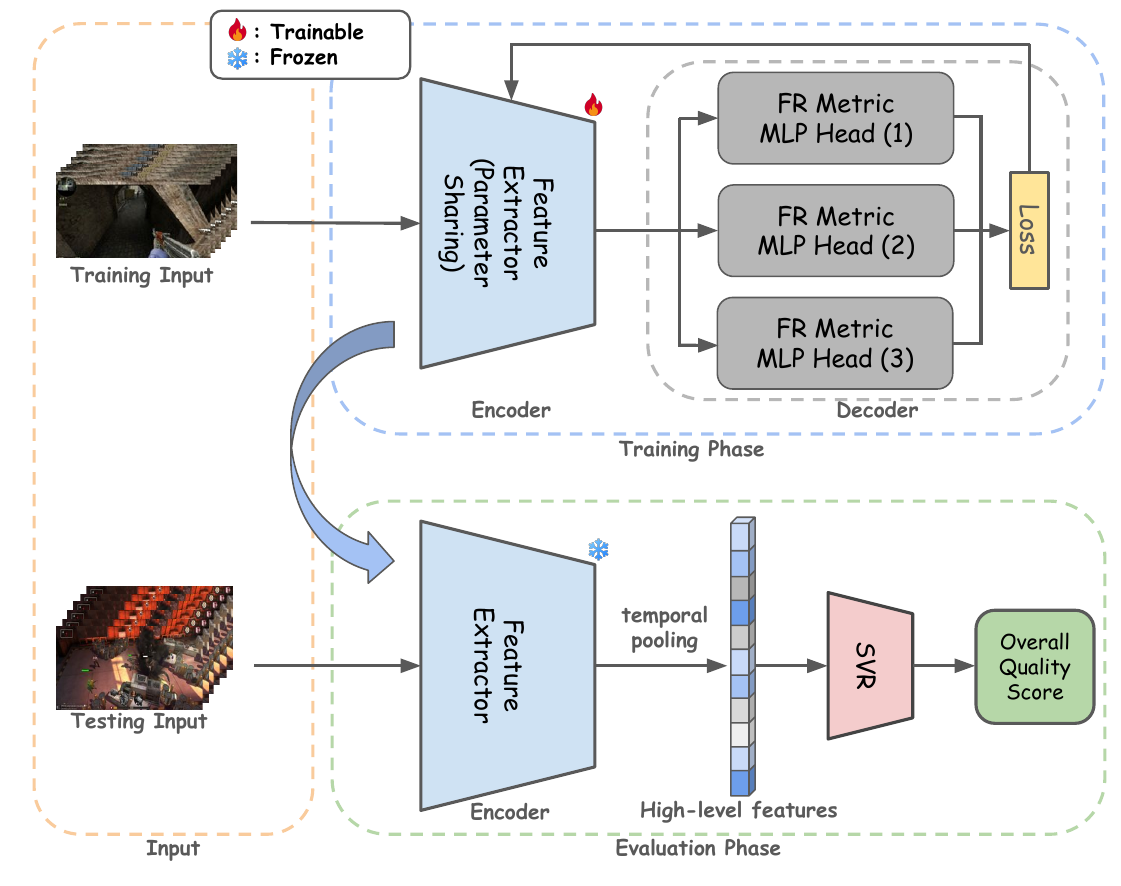}
  \caption{Overview of MTL-VQA. A shared encoder is pretrained with multiple FR objectives to learn a transferable quality representation, then frozen for downstream NR-VQA with a lightweight SVR on temporally pooled features.}
  \label{fig:framework}
\end{figure}

Despite growing practical demand, existing gaming VQA datasets, such as LIVE-YouTube Gaming~\cite{yu2023subjective}, LIVE-Meta MCG \cite{avinab2022mcg}, and the gaming subset of the YouTube UGC dataset~\cite{wang2019youtube} (hereafter referred to as YouTube UGC-Gaming), remain relatively small and sparsely labeled, limiting fully supervised deep learning-based NR models. To mitigate annotation scarcity, prior works~\cite{zadtootaghaj2018nr, goring2019nofu, barman2019no, utke2020ndnetgaming} have proposed methods to train NR predictors using a single FR proxy, such as Video Multimethod Assessment Fusion (VMAF) \cite{li2016toward}, as supervision. While effective to a degree, this strategy ties the learned representation to one specific proxy, introducing label bias and limiting generalization across domains—for example, from professionally generated content (PGC), where distortions are introduced in a controlled manner, to user-generated content (UGC), where distortions are more heterogeneous and less predictable.

At a high level, our method first learns a transferable quality representation from proxy supervision on PGC, and then adapts this frozen representation to downstream NR quality prediction on target gaming datasets using a lightweight regressor.

In this paper, we address these limitations with MTL-VQA, a multi-task framework that first learns a transferable quality representation from multiple FR proxy signals on PGC, and then adapts this frozen representation to downstream NR quality prediction on target gaming datasets using a lightweight regressor. Concretely, instead of relying on a single proxy, we treat several FR metrics as distinct supervision tasks and optimize them jointly using an adaptive gradient-balancing strategy based on the Multiple Gradient Descent Algorithm (MGDA)/MinNormSolver~\cite{sener2018multi}. After proxy-supervised pretraining on PGC, we freeze the backbone and train only a support vector regressor (SVR) on features obtained via temporal pooling, enabling label-efficient adaptation to NR-VQA on new gaming datasets.

Our contributions are summarized as follows:

\begin{itemize}[topsep=1pt,itemsep=0pt,parsep=0pt,partopsep=0pt,leftmargin=*]
\item \textbf{Label efficiency for gaming NR-VQA under domain shift.}
We show that the MTL-pretrained backbone transfers effectively under PGC-to-UGC domain shift and supports strong few-shot calibration with limited human labels. With Ridge adaptation on as few as $K{=}50$ labeled clips, MTL-VQA yields substantial gains over zero-shot transfer, and with $K{=}100$ it reaches a PLCC of \textbf{0.9301} on YouTube UGC-Gaming.

\item \textbf{Multi-proxy FR supervision with principled gradient balancing.}
We propose MTL-VQA, a multi-task pretraining framework that learns a quality-aware representation from multiple complementary FR metrics and balances them using MGDA/MinNormSolver to reduce proxy-specific dominance.

\item \textbf{Practical and reference-free deployment for cloud gaming.}
At test time, the system is fully NR and uses only a lightweight regressor on temporally pooled features from a standard ResNet-50 backbone, making it practical for real-time quality monitoring in cloud gaming. The code and pretrained models are available at \url{https://github.com/nycu-m3-lab/mtlvqa}.
\end{itemize}

\vspace{-4mm}
\section{Related Work}
\label{sec:related-work}
\vspace{-3mm}
\subsection{FR Metrics as Proxy Supervision}
\vspace{-2mm}
FR IQA/VQA estimate quality by comparing a distorted image or video against its pristine reference. Classic measures such as PSNR, SSIM~\cite{wang2004image}, and MS-SSIM~\cite{wang2003multiscale} are widely used, while learned fusion metrics such as VMAF~\cite{li2016toward} often better align with perceptual quality and have shown strong correlation with mean opinion score (MOS) in gaming streaming scenarios~\cite{barman2020objective,barman2018evaluation}. Since references are unavailable at inference in cloud gaming, FR metrics are useful as proxy supervisory signals for learning quality-aware representations and training deployable no-reference predictors.



\vspace{-3mm}
\subsection{NR-VQA Models}\label{sec:related-nr}
\vspace{-2mm}
Most NR-VQA methods were originally developed for natural content. Representative opinion-unaware handcrafted models include NIQE~\cite{mittal2012no}, BRISQUE~\cite{mittal2012making}, TLVQM~\cite{korhonen2019two}, and VIDEVAL~\cite{tu2021ugc}, while opinion-aware supervised models trained directly on human opinion scores, such as RAPIQUE\\~\cite{tu2021rapique}, VSFA~\cite{li2019quality}, FasterVQA~\cite{wu2023neighbourhood}, and DOVER++~\cite{wu2023exploring} learn directly from MOS/differential mean opinion score labels. Self-supervised quality representation learning methods, including CONTRIQUE~\cite{9796010}, ReIQA~\cite{Saha_2023_CVPR}, CONVIQT~\cite{madhusudana2023conviqt}, and CLIP-IQA+~\cite{wang2022exploring}, have also sought to improve transferability under limited labels.

Gaming videos, however, are computer-generated and contain synthetic textures, HUD overlays, and distinct motion patterns, leading to distribution shift for conventional NR models. Consequently, several gaming-oriented NR approaches have been proposed~\cite{zadtootaghaj2018nr,goring2019nofu,barman2019no}, with recent deep models leveraging higher-level semantics~\cite{utke2020ndnetgaming,chen2023gamival,yu2022perceptual}. Existing gaming NR-VQA approaches nevertheless remain limited by scarce MOS labels and frequent reliance on a single proxy objective, motivating multi-objective proxy supervision to reduce proxy-specific bias and improve transfer under limited human annotation.

\begin{figure*}[!t]
  \centering
  \includegraphics[width=500pt]{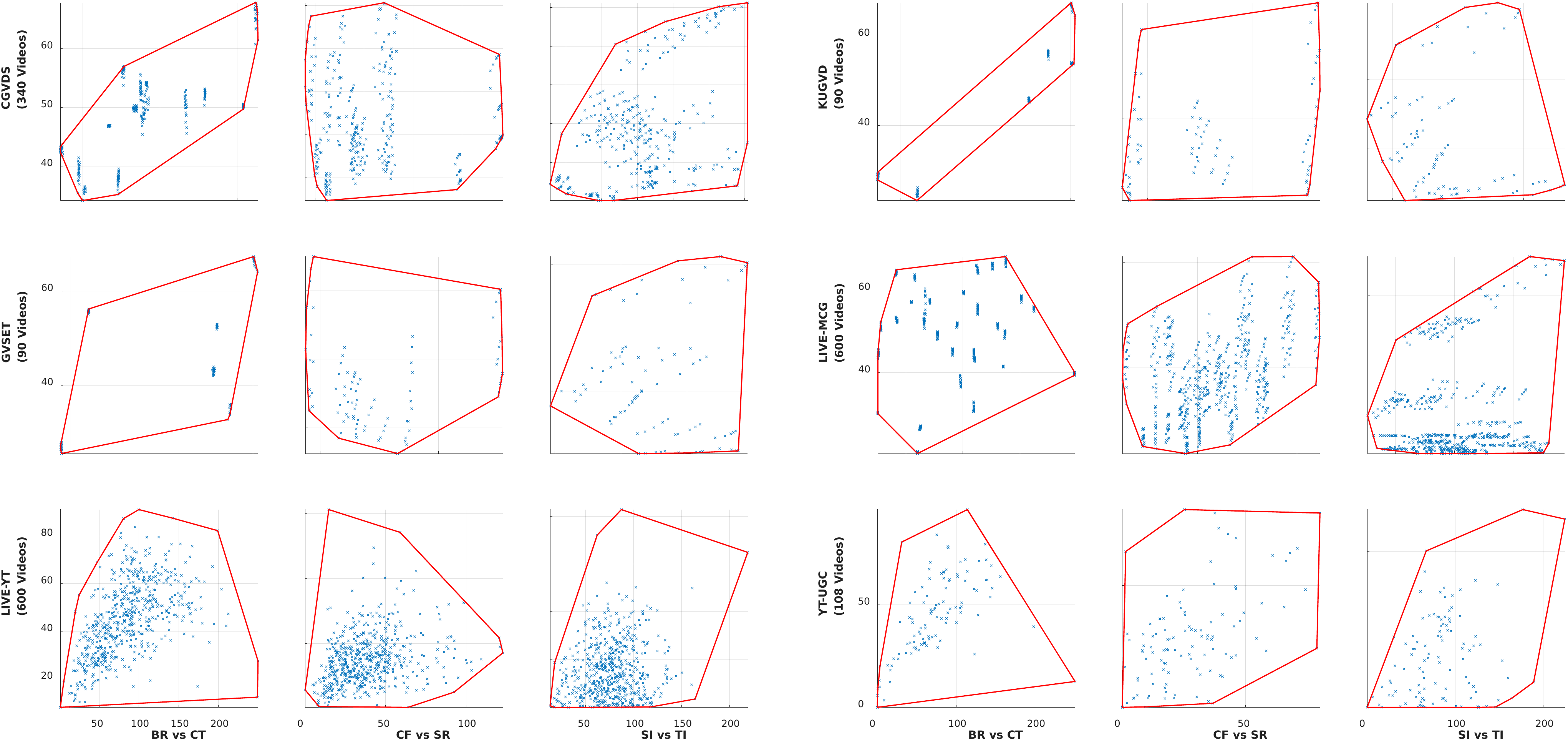}
  \caption{PGC datasets (GamingVideoSET, KUGVD, CGVDS, and LIVE-Meta MCG) and UGC datasets (LIVE-YouTube Gaming and YouTube UGC-Gaming) show different coverage patterns in the paired feature space, indicating a meaningful domain gap that motivates transferable representation learning. Blue crosses denote source samples, and red contours denote convex hulls.}
  \label{fig:domain_gap}
\end{figure*}

\vspace{-4mm}
\section{METHOD}\label{sec:format}
\vspace{-2mm}
As illustrated in Fig.~\ref{fig:framework}, our approach learns a shared representation by supervising a trainable encoder with multiple FR objectives on PGC datasets. At evaluation time, the encoder is frozen and a lightweight regressor then produces NR quality predictions on target clips. We defer training and evaluation details to the subsequent subsections. 
\vspace{-4mm}
\subsection{Problem Setup and Data}\label{sec:problem_setup}
\vspace{-2mm}
We consider gaming VQA under two operational regimes. PGC denotes controlled pipelines (e.g., cloud streaming with known encoder settings), where a pristine reference is available server-side and FR metrics can be computed on paired (reference, distorted) stimuli. UGC denotes end-user recordings and platform transcodes with heterogeneous, composite degradations; references are unavailable and human annotations are typically scarce.

\textbf{Dataset-disjoint Training/Evaluation.} Training-test partitions are dataset-disjoint by design: FR proxy supervision is derived exclusively from PGC datasets \{GamingVideoSET, KUGVD, CGVDS\} (training phase in Fig.~\ref{fig:framework}), whereas NR evaluation is conducted on separate corpora—LIVE-Meta MCG (PGC) and YouTube UGC-Gaming/LIVE-YouTube Gaming (UGC) (evaluation phase in Fig.~\ref{fig:framework}). This eliminates leakage and directly measures cross-dataset transfer.

\textbf{Proxy Target Generation (PGC).} Instead of relying only on the original reference-distorted pairs bundled in the PGC databases, we generate additional distorted streams by compressing the pristine references with \texttt{ffmpeg} under controlled bitrate ladders. Concretely, for each reference clip we encode at 0.25, 0.5, 1, 2, and 5~Mbps (other encoder parameters fixed), producing $(x, y)$ pairs at scale and enabling abundant frame supervision beyond the original database size. FR targets $y_i^{(t)}=\mathrm{FR}_t(x_i,y_i)$ including SSIM, MS-SSIM, VMAF, and FovVideoVDP~\cite{mantiuk2021fovvideovdp} are then computed between each compressed stream $x_i$ and its pristine $y_i$. Note that clip-level targets use $y^{(t)}=\frac{1}{N}\sum_i y_i^{(t)}$ when required. This procedure preserves the dataset-disjoint split (sources remain PGC) while providing rich, controllable distortions for multi-FR proxy supervision. In total, we obtain $885{,}000$ training frames for proxy supervision.

\vspace{-4mm}
\subsection{Task Formulation and Optimization}\label{sec:task_opt}
\vspace{-2mm}
Let $X=\{x_i\}_{i=1}^{N}$ and $Y=\{y_i\}_{i=1}^{N}$ be distorted/reference frames. A shared encoder $f_{\theta^{\mathrm{sh}}}$ yields $z_i=f_{\theta^{\mathrm{sh}}}(x_i)$, and $\bar z=\mathrm{Pool}(\{z_i\})$ is a clip-level representation.

\textbf{Multi-task FR Supervision.}
Each FR task $t$ attaches a lightweight head $h_{\theta^{(t)}}$  (the per-task Multilayer Perceptron [MLP] heads in Fig.~\ref{fig:framework}) to predict $\hat y_i^{(t)}$ from $z_i$. We minimize per-task regression losses over frames (or their clip average):
\[
\begin{aligned}
\mathcal{L}_t(\theta^{\mathrm{sh}},\theta^{(t)})
&= \frac{1}{N}\sum_{i=1}^{N}
   \,\phi\!\bigl(\hat y_i^{(t)},\, y_i^{(t)}\bigr), \\
\hat y_i^{(t)} 
&= h_{\theta^{(t)}}(z_i), \qquad
y_i^{(t)} = \mathrm{FR}_t(x_i, y_i).
\end{aligned}
\]

\textbf{Per-task Regression Losses.}
We use Smooth-$L_1$ loss as our loss function. Let $r=\hat y^{(t)}-y^{(t)}$. The Smooth-$L_1$ loss with parameter $\beta$ is
\[
\phi_{\text{smooth-}L_1}(r;\beta)=
\begin{cases}
\frac{r^2}{2\beta}, & |r|<\beta,\\
|r|-\frac{\beta}{2}, & \text{otherwise},
\end{cases}
\quad\text{with }\beta=1.
\]

\textbf{Adaptive Task Weighting via MinNormSolver.}\label{sec:task-opt}
To mitigate gradient interference on the shared encoder parameters $\theta^{\mathrm{sh}}$, we compute a convex combination of encoder gradients only. For each mini-batch, we first backpropagate each $\mathcal{L}_t$ separately to obtain
\(
g_t^{\mathrm{enc}}=\nabla_{\theta^{\mathrm{sh}}}\mathcal{L}_t(\theta^{\mathrm{sh}},\theta^{(t)})
\).
We then solve
\[
\alpha^\star=\arg\min_{\alpha\succeq 0,\ \|\alpha\|_1=1}
\left\| \sum_{t=1}^{T} \alpha_t\, g_t^{\mathrm{enc}} \right\|_2^2,
\]
via the geometric MinNormSolver, and form a joint loss
\(
\mathcal{L}_{\mathrm{joint}}=\sum_{t=1}^{T}\alpha_t^\star \mathcal{L}_t
\)
for the second forward/backward pass to update both $\theta^{\mathrm{sh}}$ and $\{\theta^{(t)}\}$.
\vspace{-4mm}
\subsection{Domain-Gap Considerations} 
\label{sec:domain-gap}
\vspace{-2mm}
Gaming VQA faces distribution shifts between PGC training and UGC evaluation, FR-proxy supervision and NR-MOS prediction, and varied rendering, HUD, and bitrate conditions. The separation shown in Fig.~\ref{fig:domain_gap} motivates our use of multiple complementary FR proxies during pretraining, together with zero-shot and few-shot transfer evaluation.

\begin{table}[t]
\setlength{\tabcolsep}{2pt}
\renewcommand{\arraystretch}{0.95}
\centering
\scriptsize
\resizebox{\columnwidth}{!}{%
\begin{tabular}{r|ccc|ccc|ccc}
\toprule[1.2pt]

\multicolumn{1}{r|}{\textbf{DATASET}} &
\multicolumn{3}{c|}{\textbf{GamingVideoSET}} &
\multicolumn{3}{c|}{\textbf{KUGVD}} &
\multicolumn{3}{c}{\textbf{CGVDS}} \\

\cmidrule(lr){2-4} \cmidrule(lr){5-7} \cmidrule(lr){8-10}

\textbf{Methods} &
SRCC$\uparrow$ & PLCC$\uparrow$ & RMSE$\downarrow$ &
SRCC$\uparrow$ & PLCC$\uparrow$ & RMSE$\downarrow$ &
SRCC$\uparrow$ & PLCC$\uparrow$ & RMSE$\downarrow$ \\
\midrule

PSNR      & 0.8086 & 0.8538 & 0.3959 & 0.8851 & 0.8883 & 0.3747 & 0.7301 & 0.7424 & 0.5138 \\
SSIM      & \underline{\textbf{0.9451}} & \underline{\textbf{0.9629}} & \underline{\textbf{0.2214}} 
          & \underline{\textbf{0.9603}} & \underline{\textbf{0.9659}} & \underline{\textbf{0.2097}} 
          & 0.7899 & 0.7918 & 0.4741 \\
MS-SSIM   & 0.9232 & 0.9373 & \textbf{0.2570} & \textbf{0.9541} & \textbf{0.9660} & \textbf{0.2104} & \textbf{0.8325} & \textbf{0.8310} & \textbf{0.4307} \\
VMAF      & 0.8524 & 0.8751 & 0.3953 & \textbf{0.9028} & \textbf{0.9172} & \textbf{0.3509} & \underline{\textbf{0.8894}} & \underline{\textbf{0.8969}} & \underline{\textbf{0.3417}} \\
DLM         & \textbf{0.9365} & \textbf{0.9571} & \textbf{0.2929} & 0.6591 & 0.6199 & 0.9517 & 0.6569 & 0.6298 & 0.5975 \\
VIF         & 0.9248 & \textbf{0.9462} & 0.3270 & 0.6342 & 0.5933 & 0.9763 & 0.6516 & 0.6153 & 0.6064 \\
FovVideoVDP & \textbf{0.9287} & 0.9425 & 0.3379 & 0.7114 & 0.7119 & 0.8518 & 0.5255 & 0.5729 & 0.6304 \\
LPIPS & 0.7945 & 0.8092 & 0.5272 & 0.8327 & 0.8405 & 0.5610 & 0.6454 & 0.6632 & 0.5929 \\
DISTS & 0.7977 & 0.8088 & 0.5276 & 0.8532 & 0.8628 & 0.5234 & \textbf{0.8239} & \textbf{0.8398} & \textbf{0.4301} \\
\toprule[1.2pt]
\end{tabular}
}
\caption{FR metric performance on the PGC training datasets. SSIM/MS-SSIM align best on GamingVideoSET and KUGVD, while VMAF performs best on CGVDS, motivating our default proxy subset.}
\label{tab:fr_mos_training}
\end{table}
\vspace{-4mm}
\subsection{Architecture and Implementation}
\vspace{-2mm}
Our pipeline comprises a ResNet-50 encoder $f_{\theta}(\cdot)$ (ImageNet-initialized), 
temporal pooling, and a lightweight regressor. 
Given a clip $X=\{x_i\}_{i=1}^{N}$, we extract frame embeddings 
$z_i=f_{\theta}(x_i)$ and aggregate via mean pooling to obtain 
$\bar{z}=\mathrm{Pool}(\{z_i\}_{i=1}^{N})$. 
During FR pretraining, task-specific MLP heads predict FR targets from the shared representation. 
At evaluation, the encoder is frozen and an SVR maps $\bar{z}$ to 
quality score $\hat{q}=\mathrm{SVR}(\bar{z})$, avoiding human annotations for pretraining/representation learning. Frames are downsampled (one per two) and resized to $960\times540$. 
Training uses Stochastic Gradient Descent with batch size 20. 
\vspace{-2mm}
\subsection{Multi-Task FR Supervision and Rationale}\label{sec:method-multi-task}
\vspace{-2mm}
We choose FR supervision tasks based on their empirical FR-to-MOS alignment on the PGC training datasets (Table~\ref{tab:fr_mos_training}), with emphasis on cross-dataset stability and complementarity. Structure-oriented metrics (SSIM/MS-SSIM) align best on GamingVideoSET/KUGVD, while VMAF is strongest on CGVDS. We therefore adopt VMAF, SSIM, and MS-SSIM as a compact default proxy set that balances complementary perceptual cues across datasets. To mitigate proxy-specific bias, we combine these signals via multi-task learning with adaptive weighting (Sec.~\ref{sec:task-opt}).
\vspace{-4mm}
\section{Experiments}
\vspace{-2mm}
\subsection{Experimental Setup}
\label{sec:experiments_setup}
\vspace{-2mm}
\indent\textbf{Datasets and Disjoint Training/Evaluation.}
As detailed in Sec.~\ref{sec:problem_setup}, FR proxy supervision is derived only from PGC datasets \{GamingVideoSET, KUGVD, CGVDS\}, while NR evaluation is conducted on different corpora-LIVE-Meta MCG (PGC) and two UGC testbeds (YouTube UGC-Gaming and LIVE-YouTube Gaming). This dataset-disjoint design prevents leakage and directly measures cross-dataset transfer.

For inference on the test sets, we uniformly sample 1 frame per second from each clip. Frames from LIVE-Meta MCG are resized to the dataset’s display-provided height and width, while all other evaluation datasets are resized to $960\times540$ prior to feature extraction.

\indent\textbf{Evaluation Protocols.}
We consider three protocols on each evaluation set:
(i) \textbf{Zero-shot (target-label-free; head trained on source MOS only)}: freeze the FR-pretrained encoder and train a lightweight MLP head on \emph{source} PGC datasets using their MOS labels; then freeze this head and directly apply the encoder+head to the \emph{target} test set (no target labels).
(ii) \textbf{Few-shot}: freeze the encoder and fit a radial basis function kernel SVR or a Ridge regressor on $K\!\in\!\{10,20,50,100\}$ labeled clips from the target dataset; evaluate on the remaining clips (Table~\ref{tab:few_shot_transposed}).
(iii) \textbf{Standard split}: For each evaluation dataset, we run 100 independent content-disjoint splits with an 80\%/20\% train/test ratio. The 80\% training portion is further partitioned by content into five folds for cross-validation. SVR hyperparameters $(C,\gamma)$ are selected via grid search on the training/validation folds, after which the SVR is retrained on the full training split and evaluated on the held-out 20\%. All supervised baselines and our head training follow the same split protocol.

\indent\textbf{Evaluation Metrics.}
We report rank and linear agreement with MOS using SRCC, KRCC, PLCC, and RMSE. 
Following common practice and our code path, SRCC/KRCC are computed on \emph{raw} predictions, while PLCC/RMSE are computed \emph{after} applying a logistic mapping.
Unless noted, we aggregate results over repeated content-disjoint 80/20 holdouts and report the \emph{median} across runs. 
For few-shot experiments, we additionally report medians over 100 random $K$-samplings. 

\begin{table*}[!ht]
\centering
\setlength{\tabcolsep}{2.5pt}
\renewcommand{\arraystretch}{0.95}
\scriptsize
\begin{tabular}{r|cccc|cccc|cccc} 
\toprule[1.2pt]
\multicolumn{1}{r|}{\textbf{DATASET}} & \multicolumn{4}{c|}{\makecell[c]{\textbf{LIVE-Meta MCG}\\(600 videos)}} & \multicolumn{4}{c|}{\makecell[c]{\textbf{LIVE-YouTube Gaming}\\(600 videos)}} & \multicolumn{4}{c}{\makecell[c]{\textbf{YouTube UGC-Gaming}\\(108 videos)}} \\
\cmidrule(lr){2-5} \cmidrule(lr){6-9} \cmidrule(lr){10-13}
\textbf{Methods} & SRCC$\uparrow$ & KRCC$\uparrow$ & PLCC$\uparrow$ & RMSE$\downarrow$ & SRCC$\uparrow$ & KRCC$\uparrow$ & PLCC$\uparrow$ & RMSE$\downarrow$ & SRCC$\uparrow$ & KRCC$\uparrow$ & PLCC$\uparrow$ & RMSE$\downarrow$ \\ 
\midrule[0.5pt]
NIQE & -0.3900 & -0.2795 & 0.4581 & 16.5475 & 0.2801* & N/A & 0.3037* & 16.208* & -0.5512 & -0.3889 & 0.5732 & 0.5184 \\
BRISQUE & 0.7153 & 0.5278 & 0.7206 & 12.6254 & 0.6037* & N/A & 0.6383* & 16.208* & 0.3357 & 0.2364 & 0.4588 & 0.5673 \\
TLVQM & 0.6553 & 0.4777 & 0.6889 & 13.5413 & 0.7484* & N/A & 0.7564* & 11.134* & 0.6885 & 0.5031 & 0.7267 & 0.4433 \\
VIDEVAL & 0.7601 & 0.5754 & 0.7741 & 12.0348 & 0.8071* & N/A & 0.9119* & 10.093* & 0.6736 & 0.4978 & 0.7199 & 0.4022 \\
RAPIQUE & 0.8768 & 0.7015 & 0.9043 & 8.0669 & 0.8028* & N/A & 0.8248* & 9.661* & 0.6828 & 0.5034 & 0.7399 & 0.4339 \\
GAME-VQP & 0.8773 & 0.6915 & 0.8954 & 8.3778 & \textbf{0.8563}* & N/A & 0.8754* & 8.533* & N/A & N/A & N/A & N/A \\
NDNet-Gaming & 0.8506 & 0.6619 & 0.8401 & 9.9466 & 0.4562* & N/A & 0.4690* & 14.941* & 0.2726 & 0.1871 & 0.3747 & 0.5866 \\
VSFA & 0.9143& 0.7485 & 0.9259 & 7.1455 & 0.7762* & N/A & 0.8014* & 10.396* & 0.6872 & 0.5152 & 0.7103 & 0.4806 \\
GAMIVAL & \underline{\textbf{0.9439}} & \underline{\textbf{0.7962}} & \underline{\textbf{0.9526}} & \underline{\textbf{5.6941}} & 0.8111 & N/A & 0.8321 & 9.2995 & \textbf{0.7277} & \textbf{0.5441} & \textbf{0.7640} & 0.4320 \\
CONTRIQUE & 0.9310 & 0.7768 & 0.9378 & 6.3556 & \textbf{0.8556} & \textbf{0.6738} & \textbf{0.8791} & \textbf{8.0265} & 0.7059 & 0.5306 & 0.7475 & \textbf{0.3889} \\
ReIQA & 0.7557 & 0.5669 & 0.7681 & 11.7441 & 0.3730 & 0.2609 & 0.3912 & 15.5303 & 0.5780 & 0.4163 & 0.7044 & 0.4165 \\
CONVIQT & \textbf{0.9364} & \textbf{0.7835} & \textbf{0.9423} & \textbf{6.1939} & \underline{\textbf{0.8609}} & \underline{\textbf{0.6806}} & \underline{\textbf{0.8850}} & \underline{\textbf{7.8534}} & \textbf{0.7535} & \textbf{0.5785} & \textbf{0.7894} & \textbf{0.3560} \\
CLIP-IQA+ & 0.6998 & 0.4969 & 0.7268 & 12.3443 & 0.4315 & 0.2951 & 0.5372 & 14.2693 & 0.6585 & 0.4688 & 0.6847 & 0.4611 \\
DOVER++ & 0.7855 & 0.5850 & 0.8187 & 10.3191 & 0.6530 & 0.4765 & 0.7258 & 11.6374 & 0.6822 & 0.4926 & 0.7084 & 0.4465 \\
FasterVQA & 0.4415 & 0.3046 & 0.4770 & 15.7958 & 0.2603 & 0.1767 & 0.2889 & 16.1962 & 0.2981 & 0.2019 & 0.3868 & 0.5834 \\
\midrule[0.5pt]
MTL-VQA [Ours] & \textbf{0.9434} & \textbf{0.7951} & \textbf{0.9506} & \textbf{5.7912} & 0.8486 & \textbf{0.6681} & \textbf{0.8758} & \textbf{8.1330} & \underline{\textbf{0.8292}} & \underline{\textbf{0.6468}} & \underline{\textbf{0.8643}} & \underline{\textbf{0.2924}} \\
\bottomrule[1.2pt]
\end{tabular}
\caption{Median performance over 100 random train/test splits on three gaming video quality datasets. MTL-VQA is competitive on LIVE-Meta MCG and achieves the strongest overall transfer performance on YouTube UGC-Gaming. Underlined and boldfaced entries indicate the best and top-three performers in each column, respectively. `*' denotes values cited from the authors of GAME-VQP.}
\label{tab:combined_comparison}
\end{table*}

\begin{table*}[!ht]
\centering
\renewcommand{\arraystretch}{0.95}
\renewcommand{\tabcolsep}{2.8pt}
\scriptsize
\begin{tabular}{r|cccc|ccc|cccc}
\toprule[1.2pt]
\multicolumn{1}{r|}{\textbf{DATASET}} & \multicolumn{4}{c|}{\textbf{LIVE-Meta MCG}} & \multicolumn{3}{c|}{\textbf{LIVE-YouTube Gaming}} & \multicolumn{4}{c}{\textbf{YouTube UGC-Gaming}} \\
\cmidrule(lr){2-5} \cmidrule(lr){6-8} \cmidrule(lr){9-12}
\textbf{Methods/FR combo} & SRCC$\uparrow$ & KRCC$\uparrow$ & PLCC$\uparrow$ & RMSE$\downarrow$ & SRCC$\uparrow$ & PLCC$\uparrow$ & RMSE$\downarrow$ & SRCC$\uparrow$ & KRCC$\uparrow$ & PLCC$\uparrow$ & RMSE$\downarrow$ \\
\midrule[0.5pt]
ST-VQA (VMAF only) & 0.8965 & 0.7288 & 0.9091 & 7.6649 & 0.8254 & 0.8536 & 8.7593 & 0.7377 & 0.5671 & 0.7847 & 0.3608 \\
MTL-VQA (VMAF + SSIM + MS-SSIM) & \textbf{0.9434} & \textbf{0.7951} & \textbf{0.9506} & \textbf{5.7912} & \textbf{0.8486} & \textbf{0.8758} & \textbf{8.1330} & 0.8292 & 0.6468 & 0.8643 & 0.2924 \\
\midrule[0.5pt]
VMAF + SSIM + FovVideoVDP & 0.9318 & 0.7771 & 0.9393 & 6.3611 & 0.8451 & 0.8678 & 8.3710 & 0.8631 & 0.6874 & 0.8920 & 0.2647 \\
VMAF + MS-SSIM + FovVideoVDP & 0.9315 & 0.7765 & 0.9390 & 6.3763 & 0.8447 & 0.8676 & 8.3768 & \textbf{0.8680} & \textbf{0.6918} & \textbf{0.8980} & \textbf{0.2569} \\
SSIM + MS-SSIM + FovVideoVDP & 0.9309 & 0.7762 & 0.9380 & 6.4254 & 0.8439 & 0.8690 & 8.3406 & 0.8244 & 0.6338 & 0.8652 & 0.2905 \\
\bottomrule[1.2pt]
\end{tabular}
\caption{Ablation study of proxy supervision. Compared with the single-proxy ST-VQA baseline (VMAF-only), multi-task pretraining consistently improves performance across all three benchmarks. The best three-task proxy subset is dataset-dependent: VMAF + SSIM + MS-SSIM performs best on LIVE-Meta MCG and LIVE-YouTube Gaming, whereas FovVideoVDP-containing combinations are most competitive on YouTube UGC-Gaming. We therefore use VMAF + SSIM + MS-SSIM as a strong default subset and leave adaptive proxy subset selection for future work.}
\label{tab:ablation_st_mtl}
\end{table*}

\vspace{-3mm}
\subsection{Main Results}
\vspace{-2mm}
\indent\textbf{Baselines and Results.}
We compare MTL-VQA against representative baselines from four categories discussed in Sec.~\ref{sec:related-nr}: handcrafted opinion-unaware NR-IQA/VQA models (NIQE, BRISQUE, TLVQM, VIDEVAL), opinion-aware supervised NR-VQA models (RAPIQUE, GAME-VQP, VSFA, GAMIVAL, FasterVQA, DOVER++), self-supervised quality representation learning methods (CONTRIQUE, ReIQA, CONVIQT, CLIP-IQA+), and proxy-supervised transfer methods (e.g., NDNet-Gaming).

As shown in Table~\ref{tab:combined_comparison}, MTL-VQA achieves competitive performance across all benchmarks. On LIVE-Meta MCG, it reaches 0.9434 SRCC, comparable to the best-performing baseline (GAMIVAL at 0.9439). On the more challenging YouTube UGC-Gaming testbed, MTL-VQA attains 0.8292 SRCC and surpasses strong learned baselines such as CONVIQT and DOVER++. Strong performance from contrastive-learning-based methods also highlights the importance of transferable representation learning for gaming NR-VQA under label scarcity. In comparison, our gains are achieved with multi-FR proxy supervision on PGC and a frozen backbone with a lightweight SVR head at inference, suggesting strong transfer to UGC without relying on extensive MOS-labeled training.

\indent\textbf{Few-shot Adaptation.}
Table~\ref{tab:few_shot_transposed} evaluates the label efficiency of our MTL-pretrained representations. While cross-dataset transfer without target supervision remains challenging under large gaps, the learned features provide a strong perceptual prior that can be aligned with MOS using only a small number of labeled clips. In the extreme low-data regime (e.g., $K{=}10$), Ridge regression is noticeably more stable than SVR, avoiding the numerical instability observed for SVR (e.g., undefined PLCC). On YouTube UGC, Ridge attains a PLCC of 0.7370 when the regressor is transferred across datasets ($K{=}0$) and reaches 0.9301 with only 100 labeled samples. Overall, these results highlight that multi-task FR pretraining yields transferable representations that support reliable, label-efficient calibration, which is particularly attractive for practical cloud-gaming Quality of Experience (QoE) monitoring where MOS labels are scarce.

\begin{table}[!h]
\centering
\setlength{\tabcolsep}{2.2pt} 
\renewcommand{\arraystretch}{0.95} 
\small
\resizebox{\columnwidth}{!}{%
\begin{tabular}{l|cc|cc|cc|cc}
\toprule[1.2pt]
\multirow{3}{*}{\textbf{Samples ($K$)}} & \multicolumn{4}{c|}{\textbf{LIVE-YouTube Gaming}} & \multicolumn{4}{c}{\textbf{YouTube UGC-Gaming}}\\
\cmidrule(lr){2-5} \cmidrule(lr){6-9}
& \multicolumn{2}{c|}{\textbf{SVR}} & \multicolumn{2}{c|}{\textbf{Ridge}} & \multicolumn{2}{c|}{\textbf{SVR}} & \multicolumn{2}{c}{\textbf{Ridge}} \\
\cmidrule(lr){2-3} \cmidrule(lr){4-5} \cmidrule(lr){6-7} \cmidrule(lr){8-9}
& SRCC & PLCC & SRCC & PLCC & SRCC & PLCC & SRCC & PLCC \\
\midrule[0.5pt]
$K{=}0$ & 0.2399 & 0.2228 & 0.6951 & 0.7328 & -0.1020 & 0.2821 & 0.6434 & 0.7370 \\ 
$K{=}10$ & 0.2966 & -- & 0.5114 & 0.5741 & 0.3849 & 0.3850 & 0.5937 & 0.6611 \\ 
$K{=}20$ & 0.5705 & 0.6089 & 0.6132 & 0.6745 & 0.6462 & 0.6859 & 0.6679 & 0.7273 \\
$K{=}50$ & 0.7155 & 0.7478 & 0.7158 & 0.7578 & 0.7588 & 0.8058 & 0.7449 & 0.8040 \\
$K{=}100$ & 0.7716 & 0.7996 & 0.7696 & 0.8036 & 0.7640 & 0.8867 & 0.7857 & 0.9301 \\
\bottomrule[1.2pt]
\end{tabular}
}
\caption{Zero-shot ($K{=}0$) and few-shot adaptation results of MTL-VQA on LIVE-YouTube Gaming and YouTube UGC-Gaming. For few-shot evaluation, the regressor is adapted using $K$ samples from the target dataset and evaluated on the remaining samples. Performance improves steadily as more target-domain samples are available, showing that the pretrained representation supports effective lightweight adaptation with both SVR and Ridge regressors.}
\label{tab:few_shot_transposed}
\end{table}
\vspace{-4mm}
\subsection{Ablation Studies}
\vspace{-2mm}
\indent\textbf{Single-proxy vs. multi-proxy.}
To quantify the benefit of additional proxy tasks, we compare a single-proxy baseline (ST; VMAF-only) with our multi-proxy pretraining (MTL; VMAF + SSIM + MS-SSIM). As shown in Table~\ref{tab:ablation_st_mtl}, MTL improves correlation consistently across all benchmarks. In particular, the gain is most pronounced on the YouTube UGC-Gaming benchmark, indicating that multi-FR supervision encourages a more robust and transferable perceptual representation than single-proxy one, especially across the PGC→UGC gap.

\indent\textbf{Task subset selection.}
Based on the rationale in Sec.~\ref{sec:method-multi-task}, we further evaluate alternative three-task proxy combinations in Table~\ref{tab:ablation_st_mtl}. No single combination is uniformly best across datasets: VMAF + SSIM + MS-SSIM yields the strongest results on LIVE-Meta MCG and LIVE-YouTube Gaming, whereas FovVideoVDP-containing combinations are more competitive on YouTube UGC-Gaming. This suggests that the optimal proxy subset may depend on the target distortion distribution rather than being universal. We therefore adopt VMAF + SSIM + MS-SSIM as a strong default subset and leave target-aware proxy selection, for example through a learned or validation-driven mechanism conditioned on target-domain statistics, as future work. Additional perceptual metrics such as LPIPS~\cite{Zhang_2018_CVPR} and DISTS~\cite{9298952} also remain promising candidates for extending the MTL objective.




\vspace{-4mm}
\section{Conclusion}
\vspace{-3mm}
We presented MTL-VQA, a label-efficient pretraining framework for gaming NR-VQA that leverages multiple full-reference objectives to learn robust representations. Across gaming benchmarks, a frozen backbone with a lightweight regressor achieves competitive performance, and multi-proxy supervision improves correlation over single-metric pretraining. Notably, in few-shot settings, Ridge adaptation with up to 100 labeled clips closes most of the gap to standard-split training, suggesting that representation quality, rather than regressor complexity, is the primary driver. Furthermore, from a deployment perspective, MTL-VQA supports practical cloud-gaming QoE monitoring: server-side FR proxy supervision enables a shared encoder that can be reused client-side for low-latency NR inference with minimal calibration. However, limitations remain on extremely low-quality clips where HUD overlays can dominate pooled features. To address this, future work will explore HUD-aware masking, stronger temporal modeling for fast motion and flicker, and codec-/artifact-aware auxiliary tasks to further improve robustness under diverse gaming distortions across real-world, bandwidth-constrained cloud-gaming deployments at scale.

{\footnotesize
\setlength{\itemsep}{0pt}
\setlength{\parskip}{0pt}
\bibliographystyle{IEEEbib}
\bibliography{refs}

@String(CVPR= {IEEE Conf. Comput. Vis. Pattern Recog.})

@String(TOG= {ACM Trans. Graph.})

@String(AAAI = {AAAI})

@String(CVPR  = {CVPR})

@String(TOG   = {ACM TOG})

@inproceedings{li2019quality,
  title={Quality assessment of in-the-wild videos},
  author={Li, Dingquan and Jiang, Tingting and Jiang, Ming},
  booktitle={Proceedings of the 27th ACM International Conference on Multimedia},
  pages={2351--2359},
  year={2019}
}

@ARTICLE{9796010,
  author={Madhusudana, Pavan C. and Birkbeck, Neil and Wang, Yilin and Adsumilli, Balu and Bovik, Alan C.},
  journal={IEEE Transactions on Image Processing}, 
  title={Image Quality Assessment Using Contrastive Learning}, 
  year={2022},
  volume={31},
  number={},
  pages={4149-4161},
  doi={10.1109/TIP.2022.3181496}}

@InProceedings{Saha_2023_CVPR,
    author    = {Saha, Avinab and Mishra, Sandeep and Bovik, Alan C.},
    title     = {Re-{IQA}: Unsupervised learning for image quality assessment in the wild},
  booktitle={Proceedings of the IEEE/CVF conference on computer vision and pattern recognition},
  pages={5846--5855},
  year={2023}
}

@article{barman2020objective,
  title={An objective and subjective quality assessment study of passive gaming video streaming},
  author={Barman, Nabajeet and Zadtootaghaj, Saman and Schmidt, Steven and Martini, Maria G and M{\"o}ller, Sebastian},
  journal={International Journal of Network Management},
  volume={30},
  number={3},
  pages={e2054},
  year={2020},
  publisher={Wiley Online Library}
}

@inproceedings{barman2018evaluation,
  title={An evaluation of video quality assessment metrics for passive gaming video streaming},
  author={Barman, Nabajeet and Schmidt, Steven and Zadtootaghaj, Saman and Martini, Maria G and M{\"o}ller, Sebastian},
  booktitle={Proceedings of the 23rd packet video workshop},
  pages={7--12},
  year={2018}
}

@article{mantiuk2021fovvideovdp,
  title={Fov{V}ideo{VDP}: A visible difference predictor for wide field-of-view video},
  author={Mantiuk, Rafa{\l} K and Denes, Gyorgy and Chapiro, Alexandre and Kaplanyan, Anton and Rufo, Gizem and Bachy, Romain and Lian, Trisha and Patney, Anjul},
  journal={ACM Transactions on Graphics (TOG)},
  volume={40},
  number={4},
  pages={1--19},
  year={2021},
  publisher={ACM New York, NY, USA}
}

@article{li2016toward,
  title={Toward a practical perceptual video quality metric},
  author={Li, Zhi and Aaron, Anne and Katsavounidis, Ioannis and Moorthy, Anush and Manohara, Megha},
  journal={The Netflix Tech Blog},
  volume={6},
  number={2},
  year={2016}
}

@article{mittal2012making,
  title={Making a “completely blind” image quality analyzer},
  author={Mittal, Anish and Soundararajan, Rajiv and Bovik, Alan C},
  journal={IEEE Signal Processing Letters},
  volume={20},
  number={3},
  pages={209--212},
  year={2012},
  publisher={IEEE}
}

@article{mittal2012no,
  title={No-reference image quality assessment in the spatial domain},
  author={Mittal, Anish and Moorthy, Anush Krishna and Bovik, Alan Conrad},
  journal={IEEE Transactions on Image Processing},
  volume={21},
  number={12},
  pages={4695--4708},
  year={2012},
  publisher={IEEE}
}

@article{tu2021rapique,
  title={{RAPIQUE: Rapid} and accurate video quality prediction of user generated content},
  author={Tu, Zhengzhong and Yu, Xiangxu and Wang, Yilin and Birkbeck, Neil and Adsumilli, Balu and Bovik, Alan C},
  journal={IEEE Open Journal of Signal Processing},
  volume={2},
  pages={425--440},
  year={2021},
  publisher={IEEE}
}

@article{utke2020ndnetgaming,
  title={{NDNetGaming-development of a no-reference deep CNN for gaming video quality prediction}},
  author={Utke, Markus and Zadtootaghaj, Saman and Schmidt, Steven and Bosse, Sebastian and M{\"o}ller, Sebastian},
  journal={Multimedia Tools and Applications},
  pages={1--23},
  year={2020},
  publisher={Springer}
}

@article{yu2022perceptual,
  title={Perceptual Quality Assessment of UGC Gaming Videos},
  author={Yu, Xiangxu and Tu, Zhengzhong and Birkbeck, Neil and Wang, Yilin and Adsumilli, Balu and Bovik, Alan C},
  journal={arXiv preprint arXiv:2204.00128},
  year={2022}
}

@article{chen2023gamival,
  title={{GAMIVAL}: video quality prediction on mobile cloud gaming content},
  author={Chen, Yu-Chih and Saha, Avinab and Davis, Chase and Qiu, Bo and Wang, Xiaoming and Gowda, Rahul and Katsavounidis, Ioannis and Bovik, Alan C},
  journal={IEEE Signal Processing Letters},
  volume={30},
  pages={324--328},
  year={2023},
  publisher={IEEE}
}

@article{madhusudana2023conviqt,
  title={{CONVIQT}: Contrastive video quality estimator},
  author={Madhusudana, Pavan C and Birkbeck, Neil and Wang, Yilin and Adsumilli, Balu and Bovik, Alan C},
  journal={IEEE Transactions on Image Processing},
  volume={32},
  pages={5138--5152},
  year={2023},
  publisher={IEEE}
}

@article{sener2018multi,
  title={Multi-task learning as multi-objective optimization},
  author={Sener, Ozan and Koltun, Vladlen},
  journal={Advances in neural information processing systems},
  volume={31},
  year={2018}
}

@article{korhonen2019two,
  title={Two-level approach for no-reference consumer video quality assessment},
  author={Korhonen, Jari},
  journal={IEEE Transactions on Image Processing},
  volume={28},
  number={12},
  pages={5923--5938},
  year={2019},
  publisher={IEEE}
}

@article{tu2021ugc,
  title={{UGC-VQA}: Benchmarking blind video quality assessment for user generated content},
  author={Tu, Zhengzhong and Wang, Yilin and Birkbeck, Neil and Adsumilli, Balu and Bovik, Alan C},
  journal={IEEE Transactions on Image Processing},
  volume={30},
  pages={4449--4464},
  year={2021},
  publisher={IEEE}
}

@inproceedings{zadtootaghaj2018nr,
  title={{NR-GVQM}: A no reference gaming video quality metric},
  author={Zadtootaghaj, Saman and Barman, Nabajeet and Schmidt, Steven and Martini, Maria G and M{\"o}ller, Sebastian},
  booktitle={2018 IEEE International Symposium on Multimedia (ISM)},
  pages={131--134},
  year={2018},
  organization={IEEE}
}

@inproceedings{goring2019nofu,
  title={{NOFU}—A lightweight no-reference pixel based video quality model for gaming content},
  author={G{\"o}ring, Steve and Rao, Rakesh Rao Ramachandra and Raake, Alexander},
  booktitle={2019 Eleventh International Conference on Quality of Multimedia Experience (QoMEX)},
  pages={1--6},
  year={2019},
  organization={IEEE}
}

@InProceedings{Zhang_2018_CVPR,
author = {Zhang, Richard and Isola, Phillip and Efros, Alexei A. and Shechtman, Eli and Wang, Oliver},
title = {The Unreasonable Effectiveness of Deep Features as a Perceptual Metric},
booktitle = {Proceedings of the IEEE Conference on Computer Vision and Pattern Recognition (CVPR)},
month = {June},
year = {2018}
}

@ARTICLE{9298952,
  author={Ding, Keyan and Ma, Kede and Wang, Shiqi and Simoncelli, Eero P.},
  journal={IEEE Transactions on Pattern Analysis and Machine Intelligence}, 
  title={Image Quality Assessment: Unifying Structure and Texture Similarity}, 
  year={2022},
  volume={44},
  number={5},
  pages={2567-2581},
  doi={10.1109/TPAMI.2020.3045810}}

@inproceedings{wang2019youtube,
  title={{YouTube UGC} dataset for video compression research},
  author={Wang, Yilin and Inguva, Sasi and Adsumilli, Balu},
  booktitle={2019 IEEE 21st International Workshop on Multimedia Signal Processing (MMSP)},
  pages={1--5},
  year={2019}
}

@article{avinab2022mcg,
  title={Study of subjective and objective quality assessment of mobile cloud gaming videos},
  author={Saha, Avinab and Chen, Yu-Chih and Davis, Chase and Qiu, Bo and Wang, Xiaoming and Gowda, Rahul and Katsavounidis, Ioannis and Bovik, Alan C},
  journal={IEEE Transactions on Image Processing},
  volume={32},
  pages={3295--3310},
  year={2023},
  publisher={IEEE}
}

@article{wang2004image,
  title={Image quality assessment: from error visibility to structural similarity},
  author={Wang, Zhou and Bovik, Alan C and Sheikh, Hamid R and Simoncelli, Eero P},
  journal={IEEE transactions on image processing},
  volume={13},
  number={4},
  pages={600--612},
  year={2004},
  publisher={IEEE}
}

@inproceedings{wang2003multiscale,
  title={Multiscale structural similarity for image quality assessment},
  author={Wang, Zhou and Simoncelli, Eero P and Bovik, Alan C},
  booktitle={The thrity-seventh asilomar conference on signals, systems \& computers, 2003},
  volume={2},
  pages={1398--1402},
  year={2003},
  organization={Ieee}
}

@article{barman2019no,
  title={No-reference video quality estimation based on machine learning for passive gaming video streaming applications},
  author={Barman, Nabajeet and Jammeh, Emmanuel and Ghorashi, Seyed Ali and Martini, Maria G},
  journal={IEEE Access},
  volume={7},
  pages={74511--74527},
  year={2019},
  publisher={IEEE}
}

@article{yu2023subjective,
  title={Subjective and Objective Analysis of Streamed Gaming Videos},
  author={Yu, Xiangxu and Ying, Zhenqiang and Birkbeck, Neil and Wang, Yilin and Adsumilli, Balu and Bovik, Alan C},
  journal={IEEE Transactions on Games},
  volume={16},
  number={2},
  pages={445--458},
  year={2023},
  publisher={IEEE}
}

@article{wu2023neighbourhood,
  title={Neighbourhood representative sampling for efficient end-to-end video quality assessment},
  author={Wu, Haoning and Chen, Chaofeng and Liao, Liang and Hou, Jingwen and Sun, Wenxiu and Yan, Qiong and Gu, Jinwei and Lin, Weisi},
  journal={IEEE Transactions on Pattern Analysis and Machine Intelligence},
  volume={45},
  number={12},
  pages={15185--15202},
  year={2023},
  publisher={IEEE}
}

@inproceedings{wang2022exploring,
    author = {Wang, Jianyi and Chan, Kelvin CK and Loy, Chen Change},
    title = {Exploring CLIP for Assessing the Look and Feel of Images},
    booktitle = {AAAI},
    year = {2023}
}

@inproceedings{wu2023exploring,
  title={Exploring video quality assessment on user generated contents from aesthetic and technical perspectives},
  author={Wu, Haoning and Zhang, Erli and Liao, Liang and Chen, Chaofeng and Hou, Jingwen and Wang, Annan and Sun, Wenxiu and Yan, Qiong and Lin, Weisi},
  booktitle={Proceedings of the IEEE/CVF International Conference on Computer Vision},
  pages={20144--20154},
  year={2023}
}
}

\end{document}